\documentclass[aps,prl,showpacs,twocolumn,superscriptaddress,letterpaper]{revtex4}

\usepackage{graphicx}
\usepackage{dcolumn}
\usepackage{bm}
\usepackage{amsmath, amssymb}%
\usepackage{epstopdf}
\usepackage{color}
\usepackage[normalem]{ulem}

\begin{document}



\title{Analysis of Neutron Stars Observations Using a Correlated Fermi Gas Model}

\newcommand*{\MIT }{Massachusets Institute of Technology, Cambridge, MA 02139, USA}
\newcommand*{\MITindex}{1}
\affiliation{\MIT} 
\newcommand*{\INT }{Institute for Nuclear Theory, University of Washington, Seattle, Washington 98195, USA}
\newcommand*{\INTindex}{2}
\affiliation{\INT} 
\newcommand*{\Tennessee }{Department of Physics and Astronomy, University of 
Tennessee, Knoxville, Tennessee 37996, USA}
\newcommand*{\Tennesseeindex}{3}
\affiliation{\Tennessee} 
\newcommand*{\ORNL }{Physics Division, Oak Ridge National Laboratory, Oak Ridge, Tennessee 37831, USA}
\newcommand*{\ORNLindex}{4}
\affiliation{\ORNL} 
\newcommand*{\TAU }{Tel Aviv University, Tel Aviv 69978, Israel}
\newcommand*{\TAUindex}{5}
\affiliation{\TAU} 
\newcommand*{\ODU }{ Old Dominion University, Norfolk, VA 23529, USA} 
\newcommand*{\ODUindex}{6}
\affiliation{\ODU }

  
\author{O. Hen}
     \affiliation{\MIT}
     \email[Contact Author \ ]{hen@mit.edu}
\author{A.W. Steiner}
     \affiliation{\INT}
     \affiliation{\Tennessee}
     \affiliation{\ORNL}
\author{E. Piasetzky}
     \affiliation{\TAU}
\author{L.B.~Weinstein}
     \affiliation{\ODU}

\date{\today}

\begin{abstract}

  {\bf Background:} The nuclear symmetry energy is a fundamental
  ingredient in determining the equation of state (EOS) of neutron
  stars (NS).  Recent terrestrial experiments constrain both its value and
  slope at nuclear saturation density, however, its value at
  higher densities is unknown.  Assuming a Free Fermi-gas (FFG) model
  for the kinetic symmetry energy, the high-density extrapolation
  depends on a single parameter, the density dependence of the
  potential symmetry energy.  The Correlated Fermi-gas (CFG) model
  improves on the FFG model by including the effects of short-range,
  correlated, high-momentum, nucleons in nuclear matter.  Using the
  CFG model for the kinetic symmetry energy along with constraints
  from terrestrial meaurements leads to a much softer density
  dependence for the potential symmetry energy.

  {\bf Purpose:} Examine the ability of the FFG and CFG models to
  describe NS observables that are directly sensitive to the
  symmetry energy at high-density. Specifically, examine the ability
  of the CFG model, with its softer density dependence of the
  potential symmetry energy, to describe a two solar-mass NS.

{\bf Methods:} Using a Bayesian analysis of NS
observables, we use the CFG and FFG models to describe the symmetry
energy and examine the resulting parameters in the NS EOS
and the density dependence of the potential symmetry energy.

{\bf Results:} Despite the large difference in the density dependence
of the potential part of the symmetry energy, both models can
describe the NS data and support a two solar-mass NS. The different
density dependences has only a small effect on the NS EOS.

{\bf Conclusions:} While sensitive to the high-density values of the
symmetry energy, NS observables alone are not enough to
distinguish between the CFG and FFG models.  This indicates that the
NS EOS, obtain from Bayesian analysis of NS observables, is
robust and is not sensitive to the exact nuclear model used for the
kinetic part of the nuclear symmetry energy.

\end{abstract}

\pacs{
}

\maketitle

%


\section{Introduction}
Determining the equation of state (EOS) of dense nuclear matter, such
as that found in neutron stars (NS), is a long-sought goal of
nuclear physics. The EOS is a fundamental property
  of quantum chromodynamics and is an observable independent
  of renormalization scale and scheme. While considerable progress had been made
in theoretical studies of nuclear and neutron matter at high
densities~\cite{Rios:2013zqa,Gandolfi:2011xu,Hebeler:2010jx},
experimental constraints from terrestrial experiments and
astrophysical observations are still sparse.

One of the largest uncertainties in the NS EOS is the density dependence of the nuclear symmetry energy~\cite{Prakash:1988md}. 
This describes the change in the energy of nuclear matter as one replaces a proton with a neutron. 
The symmetry energy is constrained by terrestrial measurements up to
nuclear saturation density, $\rho_0$ (= 0.16 nucleons/fm$^3$
$\approx$ 160
MeV/fm$^3$)~\cite{Lattimer:2014sga,Li:2013ola,Lynch:2009vc,Trautmann:2012nk,Tsang:2012se,Li:2014oda,Horowitz:2014bja,Lattimer:2012nd}.
Specifically at saturation density, the value and slope of the
symmetry energy were recently determined to within an accuracy of about $\pm3$ MeV and $\pm20$ MeV respectively~\cite{Lattimer:2014sga,Li:2013ola}.
The symmetry energy behavior at supra-nuclear densities, required for the description of NS, is not well known.

A common method used to simplify the extraction of the density
dependence of the symmetry energy is to split the symmetry energy into
kinetic and potential parts and study them
separately~\cite{Tsang:2008fd}.  The kinetic term is usually
determined analytically using a zero-temperature Free-Fermi Gas (FFG)
model, which fully determines the value at saturation density and the
density dependence to supra-nuclear densities.  Combined with the
known total symmetry energy at saturation density, this determines the
potential symmetry energy at saturation density, leaving its density
dependence as the only unknown~\cite{Tsang:2008fd,Steiner:2010fz}.

While the analytical FFG model is simple and easy to use, it fails to
describe many relevant properties of nuclear systems.  In particular,
microscopic calculations have shown that the FFG model underestimates
the kinetic energy carried by nucleons in nuclei and nuclear
matter~\cite{wiringa14,cda96,Rios:2013zqa,Carbone12,Vidana11,Lovato:2010ef}.
In addition, the FFG parameterization fails to accurately describe
quantum Monte Carlo calculations of pure neutron matter~\cite{Gandolfi:2011xu}.
Results from recent electron-scattering experiments indicate that $20$
to $25\%$ of nucleons in medium and heavy nuclei have momentum greater
than the Fermi momentum~\cite{egiyan03,egiyan06,fomin12}.  These
high-momentum nucleons dominate the kinetic energy of nucleons in
nuclei and are predominantly in the form of neutron-proton (np)
short-range correlated (SRC)
pairs~\cite{piasetzky06,tang03,shneor07,subedi08,korover14,hen14}.
These SRC pairs are pairs of nucleons with large relative momentum and
small center-of-mass momentum, where large and small are relative to
the Fermi momentum.  By neglecting these SRC pairs, the FFG model
significantly underestimates the kinetic energy carried by nucleons in
nuclei~\cite{hen15}.

The effect of np-SRC pairs on the nuclear symmetry energy was recently
investigated using the Correlated Fermi-Gas (CFG) model~\cite{hen15}.
This model describes the momentum distribution of nucleons in symmetric
nuclear-matter by:
\begin{equation}
n^{SRC}_{SNM}(k)=\left\{ \begin{array}{ll}
A_0 & k<k_F\\
C_{\infty}/k^4 & k_F<k<\lambda k^0_F ,\\
0 & k>\lambda k^0_F
\end{array}
\right. 
\label{eq:nk}
\end{equation}
where $A_0$ describes a depleted Fermi gas distribution that extends
up to $k_F$, the density dependent Fermi momentum. Above the Fermi
momentum the momentum distribution is dominated by np-SRC pairs and
falls off as $C_{\infty}/k^4$~\cite{Hen:2014lia}.  This high momentum
tail extends from $k_F$ to a constant cutoff given by $\lambda k^0_F$,
where $k^0_F$ is the Fermi-momentum of symmetric nuclear matter at
saturation density.  All constants in Eq.~\ref{eq:nk} (e.g.,
$C_{\infty}=c_0k_F$ where $c_0=4.16\pm0.95$ and $\lambda=2.75\pm0.25$)
are extracted from data, see Ref.~\cite{hen15} for details.

The kinetic symmetry energy calculated using the FFG and the CFG
models differ significantly.  At saturation density, the FFG kinetic symmetry energy is
either 12.5 or 17.0 MeV and the CFG kinetic symmetry energy ranges
from $-2.5$ to $-17.5$ MeV~\cite{hen15}.  The density
dependences of the kinetic symmetry energy also differ.  The FFG
kinetic symmetry energy is proportional to $\rho^{2/3}$ while the CFG
kinetic symmetry has terms proportional to $\rho^{1/3}$, $\rho^{2/3}$, and
$\rho$.  

The total symmetry energy equals the sum of the kinetic and potential
symmetry energies. Since the value
($29~\mathrm{MeV}<E_{sym}(\rho_0)<36~\mathrm{MeV}$) and density
dependence ($30~\mathrm{MeV}<L=3\rho\frac{\partial
  E_{sym}(\rho)}{\partial\rho}\vert_{\rho_0} <70~\mathrm{MeV}$) of the
total symmetry energy are known at saturation
density~\cite{Lattimer:2014sga}, the value and density dependence of
the potential symmetry energy will depend on the model used for the
kinetic symmetry energy. 

This work examines the sensitivity of the NS EOS, extracted from
Bayesian analysis of NS mass and radius
observations~\cite{Steiner:2010fz}, to the inclusion of np-SRC using
the CFG model.  This is a complementary and independent approach to
the previous use of terrestrial observations at saturation density and
has a larger sensitivity to the high-density behavior of the symmetry
energy.  

We start with a short overview of NS observables, EOS
parameterizations, and Bayesian analysis used to constrain free
parameters in the NS EOS. We then discuss our results with emphasis on
similarities and differences in the NS EOS obtained using the FFG and
CFG models.  We highlight the robustness of the resulting EOS and
discuss the differences in the extracted potential symmetry energy.

\begin{figure} [b]
\includegraphics[width=8cm]{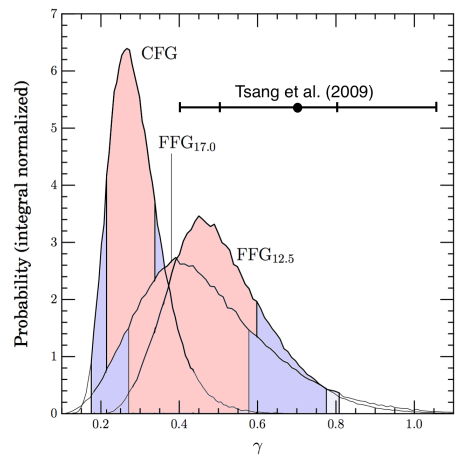}
\caption{\label{fig:gamma} (color online) The probability distribution
  of the extracted potential symmetry energy density dependence
  parameter $\gamma$ (detailed in Eq.~\ref{eq:2}), obtained from a
  Bayesian analysis of NS observations using the CFG, FFG$_{12.5}$,
  and FFG$_{17.0}$ models for the kinetic symmetry energy. The inner
  and outer shaded region mark the $1$- and $2$-$\sigma$ limits of
  each distribution, see text for details.  The horizontal line shows
  the centroid and one- and two-sigma limits from an analysis of heavy-ion
  collision data \cite{Tsang:2008fd}.}
\end{figure}

\begin{figure*}[t]
\includegraphics[width=6cm]{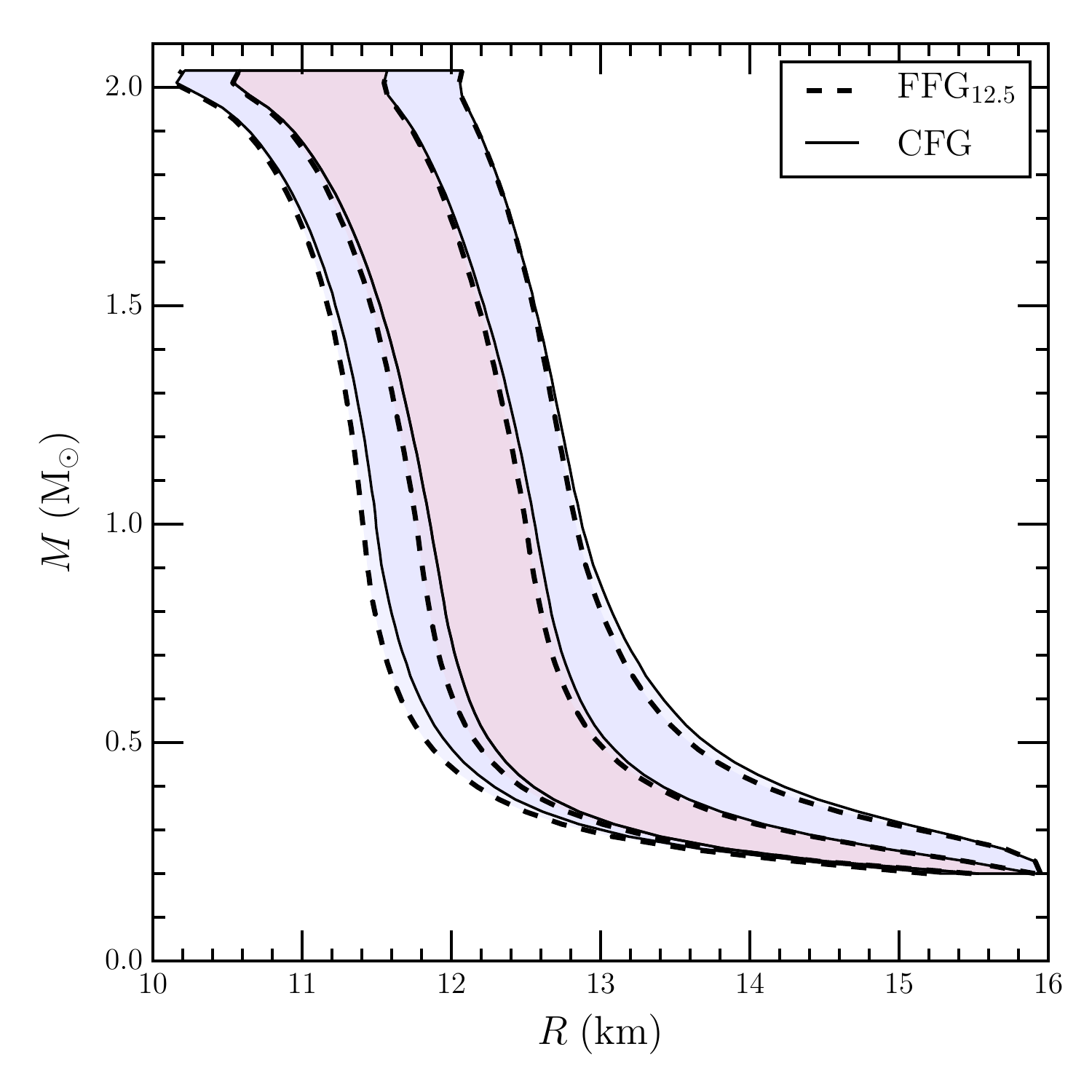}
\includegraphics[width=6cm]{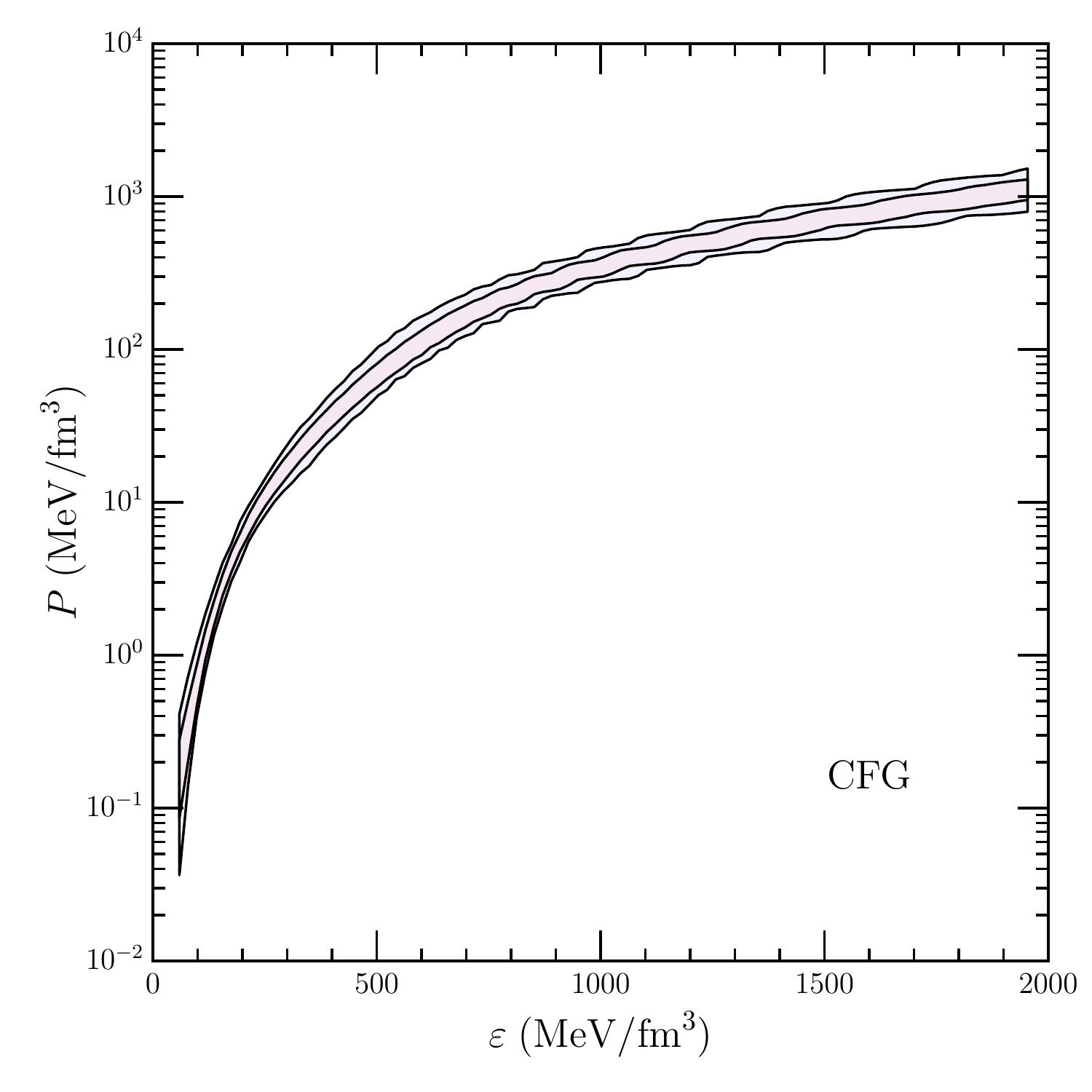}
\caption{The extracted mass-radius (left) and pressure energy-density
  (right) relations for the CFG (solid line) and FFG$_{12.5}$ (dashed
  line) models.  The results
  for the FFG$_{12.5}$ and FFG$_{17}$ models are almost identical.
  The inner and outer contours show the one- and two-sigma results.}
\label{fig:EOS}
\end{figure*}

\section{ Bayesian analysis of NS observables and the NS EOS} 
Bayesian analysis allows constraining the NS EOS by performing a
global fit of NS EOS to NS mass-radius extractions, taking into
account external constraints from terrestrial measurements,
astrophysical observations (e.g., observation of a two solar-mass NS)
and physical limitations such as causality (i.e., speed of sound $\le$
speed of light), and hydrodynamical
stability~\cite{Steiner:2010fz,Steiner:2012xt}.

The NS observations used in the analysis presented here include high
precision mass extractions from Pulsar-timing measurements,
simultaneous mass-radius extractions from photospheric radius
expansion (PRE) X-ray burst measurements, and thermal spectra
measurement of low-mass X-ray Binaries (LMXB), see
Ref.~\cite{Steiner:2012xt} for details.

The parameterization of the NS EOS is divided into three
energy-density regions: low ($\le15$ MeV/fm$^3$), medium ($15$ to
$\approx350$ MeV/fm$^3$), and high ($\le\approx350$ MeV/fm$^3$).  The
low energy-density region describes the NS crust and its functional
form is assumed to be well constrained.  The high energy-density
region is parameterized by a one or two polytropes.  The medium
energy-density region has a physically motivated functional form, with
two fit parameters (Incompressibility, $K$, and Skewness, $\kappa$)
and the density dependent symmetry energy.  See
Ref.~\cite{Steiner:2010fz} for details.

As described in the introduction, the total symmetry energy is generally given by:
\begin{equation}
E_{sym}(\rho/\rho_0)=E_{sym}^{kin}(\rho/\rho_0)+E_{sym}^{pot}(\rho/\rho_0),
\label{eq:1}
\end{equation}
where $E_{sym}^{kin}(\rho/\rho_0)$ and $E_{sym}^{pot}(\rho/\rho_0)$
are the kinetic and potential parts of the total symmetry energy.  At
nuclear saturation density the total symmetry energy, $S_v$, and its
slope, $L$, are well constrained by terrestrial
measurements~\cite{Lattimer:2014sga,Li:2013ola}.  The kinetic term can
be analytically calculated assuming a FFG or CFG model, and the
potential symmetry energy at saturation energy is calculated as:
$S_{pot}$ = $S_v - S_{kin}$, where $S_{kin}$ is the kinetic symmetry
energy at saturation density.  The density dependence of the potential
symmetry energy is parameterized as:
\begin{align}
E_{sym}^{pot}(\rho/\rho_0) &  =S_{pot}\cdot(\rho/\rho_0)^{\gamma} \nonumber \\
                                                 & =(S_{v}-S_{kin})\cdot(\rho/\rho_0)^{\gamma},
\label{eq:2}
\end{align}
where, assuming knowledge of $S_{kin}$, $\gamma$ is the only unknown. 

To constrain the NS EOS in a self-consistent way, we follow Steiner et
al.~\cite{Steiner:2010fz} and perform a Bayesian analysis of all
available NS observations and terrestrial constraints on $S$ and $L$, using the FFG
or the CFG models to express the kinetic symmetry energy at saturation
and its density dependence. Previous studies used the FFG model with
either a free nucleon or an effective nucleon mass, resulting in
$S_{kin}$ = $12.5$ and $17.0$ MeV respectively \cite{Tsang:2008fd,Steiner:2010fz}. We examine both
options and refer to them as FFG$_{12.5}$ and FFG$_{17.0}$
respectively.

\begin{figure*}[t]
\begin{center}
\includegraphics[width=6cm]{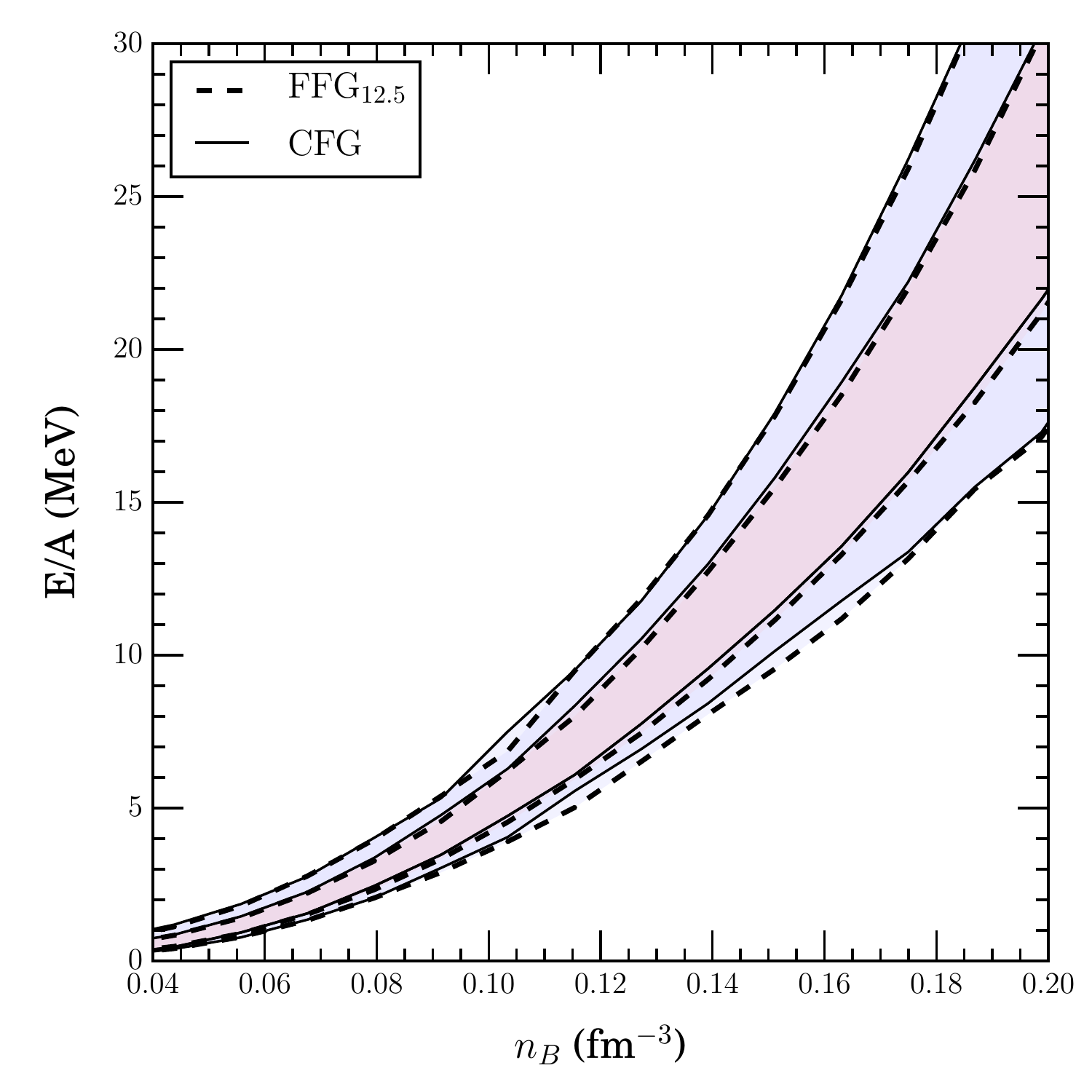}
\caption{\label{fig:Enegy} (color online) The extracted energy per
  particle as a function of the baryonic density for the CFG (solid lines)
  and FFG$_{12.5}$ (dashed lines) models. The results of the FFG$_{12.5}$
  and FFG$_{17}$ models are almost identical.  The inner and outer
  contours show the one- and two-sigma limits.  Both models are consistent with the 
  empirical value of 16 MeV at saturation density.}
\end{center}
\end{figure*}

\section{Bayesian Analysis Results}
We start by examining the details of the potential symmetry
energy. Fig.~\ref{fig:gamma} shows the density dependence of the
potential symmetry energy for the three models. As can be seen, this variable is
very sensitive to the choice of the kinetic symmetry energy model. The
CFG kinetic symmetry energy is significantly lower than that of the
FFG at saturation density. Because the total symmetry energy and its
slope are fixed at saturation density, this increases the potential
symmetry energy at $\rho_{0}$ and drastically decreases its density
dependence, $\gamma$.

We note that the results shown in Fig.~\ref{fig:gamma} for the
FFG$_{17.0}$ model differ from the ones previously obtained from a
similar Bayesian analysis using the FFG$_{17.0}$
model~\cite{Steiner:2010fz}.  This difference is due to the inclusion of
additional observations in the analysis described here
and the expansion of the allowed range for $\gamma$ down to zero 
to allow a clear comparison with the CFG model.  Unlike previous works,
the density dependence ($\gamma$) obtained using the FFG models are consistent with that
extracted from heavy-ion analysis using the FFG$_{12.5}$
model~\cite{Tsang:2008fd}.

The dramatically different potential symmetry energy and density
dependence obtained using the CFG and FFG models does not appear to
have a large effect on the bulk properties of the resulting NS EOS.
Fig.~\ref{fig:EOS} shows the EOS obtained from the Bayesian analysis
using the CFG and FFG$_{12.5}$ models (the 
FFG$_{12.5}$ and FFG$_{17.0}$ models give almost identical results).  Notice that despite the soft
density dependence of the potential symmetry energy, the CFG
EOS supports a two solar-mass NS.  

Fig.~\ref{fig:Enegy} shows the
extracted energy per nucleon as a function of the baryonic density for
the CFG and FFG$_{12.5}$ models (results for the FFG$_{17.0}$ model
are practically identical to the FFG$_{12.5}$ model). The FFG results
here are also very similar to the CFG model, although the latter
yields a slightly larger energy. Both models are consistent with the
empirical value of $E/A = 16$ MeV at saturation density.

The almost identical EOS and energy-density relations for the CFG,
FFG$_{12.5}$ and FFG$_{17.0}$ models (as shown in Figs.~\ref{fig:EOS}
and~\ref{fig:Enegy}) support the robustness of the Bayesian analysis
and indicates that it is insensitive to the exact nuclear model used
for the kinetic term of the nuclear symmetry energy. This is not
surprising, since these are bulk properties of nuclear matter, which
depend on the sum of the kinetic and potential symmetry energies
(which is the same for both the CFG and FFG models).

The bulk properties of NS are robust and largely insensitive to the
choice of the kinetic symmetry energy model.  However, it is desired
to know (1) which model captures the nuclear dynamics better and (2)
whether there
are other observables that can differentiate between them?
Recent calculations done using a Relativistic Mean-Field (RMF) model
for the symmetry potential obtained very different results for the
nuclear incompressibility when calculated using the CFG and FFG
models~\cite{Cai:2015xga}.
The result of the CFG model was consistent with recent
experimental constraints. Another possible test could come from pion
production and isospin diffusion observables measured in
intermedium-energy heavy-ion collisions. These observables are
directly sensitive to the potential symmetry energy but are
traditionally analyzed using transport models that only incorporate
the FFG model. By incorporating SRCs into transport models one could
possibly differentiate between the FFG and CFG models.

\section{Summary} 
The kinetic part of the nuclear symmetry energy can be parametrized
using two models: CFG and FFG. The CFG model includes short-range high-momentum pairs of nucleons in
nuclei; the FFG model does not. Using Bayesian analysis of NS observables, we
examined the ability of the CFG and FFG models to describe the data
and examined the resulting parameters in the NS EOS and the
density dependence of the potential symmetry energy. We find that both
models can describe the data and support a two solar-mass NS. The
obtained density dependence for the potential part of the symmetry
energy is very different between the two models, but this has a small
effect on the NS EOS.

While sensitive to the high-density values of the symmetry energy,
NS observables alone cannot distinguish between
the CFG and FFG models.  This indicates that the NS EOS, obtained from
Bayesian analysis of NS observables is robust and is not sensitive to the exact
nuclear model used for the kinetic term of the nuclear symmetry 
energy.

\begin{acknowledgments}
We thank Bao-An Li and Misak Sargsian for many fruitful discussions.
This work was partially supported by the U.S. Department of Energy
under grant No. DE-SC00006801, 
the Israel Science Foundation under grant No. DE-FG02-96ER40960,
and by the National Science Foundation under grant PHY 1554876.
\end{acknowledgments}

\bibliography{eep}

\end{document}